\documentclass[preprint2]{aastex}

\begin{document}

\title{Magnetospheric Scattering and Emission in Millisecond Pulsars}

\author{Timothy M. Braje and Roger W. Romani}
\affil{Department of Physics, Stanford University,
    Stanford, CA 94035}
\email{timb@astro.stanford.edu;rwr@astro.stanford.edu}

\begin{abstract}

We model the formation of magnetospheric components of millisecond pulsar
light curves, deriving an approximate model for the curved space,
`swept-back' dipole field and following photon emission, propagation,
and scattering.  Magnetospheric pulse components are strongly affected
by rapid rotation and Schwarzschild effects.

\end{abstract}


\section{Introduction and Light Curve Model}

With the launch of {\it CXO} and {\it XMM-Newton}, there has been a
dramatic increase in sensitivity for high resolution X-ray studies of
compact objects.  Rapidly rotating neutron stars are targets of
particular interest, since measurements of the stellar surface can
probe the neutron star evolution and the equation of state at high
density.  Several relativistic effects may also be visible, in some
cases.  A number of millisecond pulsars (MSP) have been observed in
the X-rays, both as isolated rotation powered objects \citep{bec99}
and as accretion powered LMXBs \citep[e.g.][]{wij98}.  In
\citet*{bra00}, we explored the distortions of the surface emission
imparted by rapid rotation, including the important effects of Doppler
boosting, aberration, and gravitational focusing and time delays. More
subtle effects induced by frame dragging were also considered.  For
rotation-powered pulsars, the surface radiation must traverse the
magnetosphere, where resonant scattering can introduce significant
distortions in the pulse profile \citep{raj97}.  Further, several
X-ray MSP have narrow, non-thermal pulse components, suggesting a
direct origin in the magnetosphere.  In this paper, we extend
the treatment of rapid rotation effects to include scattering and
emission in a surrounding dipole magnetosphere.

Our model assumes a spherical star, which we take to be sufficiently
centrally condensed that a Schwarzschild (or Kerr) metric is an
adequate model of the external spacetime (the `Roche
approximation'). Because we wish to extend the modeling to small spin
periods $P_*$, the effects of `sweep-back' on the magnetic field, even
for $r$ a few times the stellar mass $M$, can be substantial.  In
addition, the rapid rotation develops large electro-motive forces.  We
follow the common assumption that this EMF causes pair production such
that the closed zone of the magnetosphere is filled with a
charge-separated pair plasma, whose charge distribution cancels the
rotational EMF.  For the magnetic field, we assume a point dipole
located at the stellar center and generalize the computation of the
field structure for the non-rotating dipole in the Schwarzschild
spacetime \citep[e.g.,][]{bar73,pra97,mus97}.  Our result passes
smoothly to the Schwarzschild magnetic field {\boldmath $B$} as the
stellar angular velocity $\Omega_*$ vanishes and to the flat space
rotating dipole solution as $r \rightarrow \infty$.  We assume that
the plasma co-rotates (is stationary) in the closed zone, which is
defined by the `last closed' field lines traced from tangent approach
to  the light cylinder at $r_{\rm LC}=c\,P_*/2\pi$ to the stellar
surface.

Our light curve modeling follows the procedure described in
\citet{bra00}.  This Monte Carlo code starts with photons randomly
drawn from the surface emission zone, with initial directions drawn
from a model limb-darkened distribution.  We aberrate these surface
photons and propagate to infinity through curved space, using the
Schwarzschild or Kerr metric, as appropriate.  As the photon passes
through the magnetosphere, we monitor for local cyclotron resonance.
At the resonance position photons are re-emitted with the proper
boosted scattering angular distribution.  We also follow photons
arising directly from the magnetosphere --- from acceleration gaps or
other non-thermal sources.  After including the gravitational and
time-of-flight delays, the photons are assigned to energy and
rotational phase bins to produce maps of the radiation on the sky.
Slices through these maps at the Earth's viewing angle provide pulsar
light curves and phase-resolved spectra.

\section{Magnetic Field Structure}

\begin{figure*}[tb]
\epsscale{2.1}
\plotone{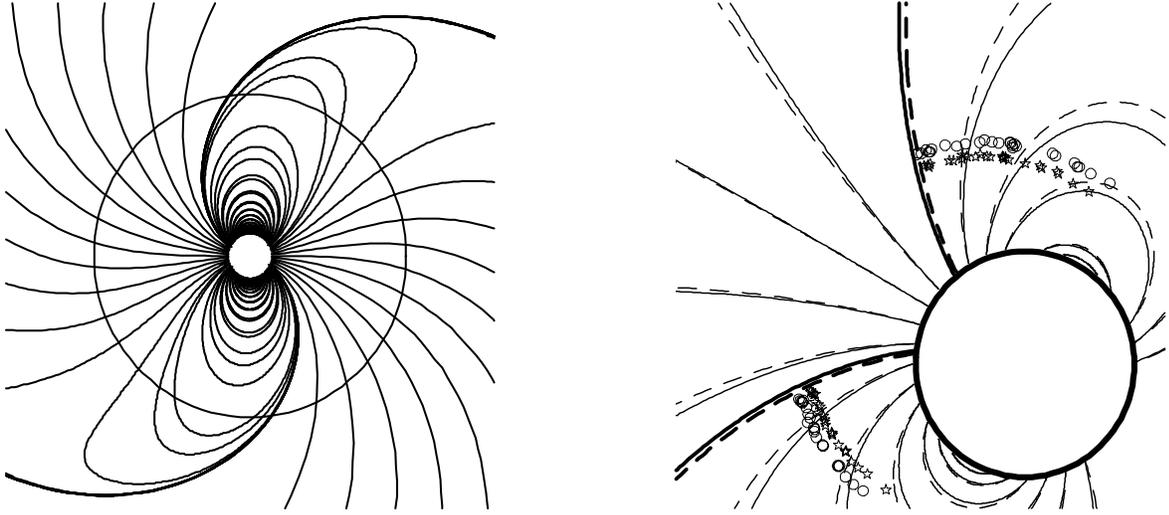}
\caption{
\emph{Left Panel} -- Retarded dipolar magnetic field viewed down the rotational axis.
The light cylinder is depicted as a circle on top of the field lines. 
\emph{Right Panel} -- Curved space magnetic field perturbations viewed in
the {\boldmath $\mu$}-{\boldmath $\Omega_*$} plane.  Flat space (dashed lines)
and curved space (solid lines) --- the `last closed' field 
lines are drawn bolder.  The points show the closed zone resonant scattering 
locus for Schwarzschild (circles) and flat space (stars) fields, $P_*=1.5$ms, 
and the same narrow range of photon energy (2.8eV--3.0eV).  $B$ is set by the 
same magnetic moment $\mu(r=\infty)$.
\label{magfld}}
\end{figure*}

We have developed an approximate, retarded, dipolar magnetic field
expression that links the exact non-rotating Schwarzschild result at
small $r$ with the flat space `swept-back' field structure at large
$r$.  We expect the magnetic field
lines to have approximately the same shape as a static dipole near the
star, but to curve back near the light cylinder, as Figure~\ref{magfld}
displays.

To derive our magnetic field expression, we write Maxwell's Stress
Tensor as derivatives of the vector potential in covariant form as
\begin{equation}
	F_{\mu\nu} = \frac{\partial A_{\nu}}{\partial x^{\mu}} -
			\frac{\partial A_{\mu}}{\partial x^{\nu}}. \label{maxstr}
\end{equation}
We use the Schwarzschild metric in geometrized units ($G=c=1$):
\begin{equation}
	{\rm d}s^{2}= \eta^2 {\rm d}t^{2} - \frac{ {\rm d}r^{2}}{\eta^2}
 		-r^{2} {\rm d}\theta^{2} - r^{2} \sin^{2} \theta {\rm d}\phi^{2}
\end{equation}
where $\eta = \sqrt{1-2M/r}$.  If we solve Maxwell's Equations in vacuum
\begin{equation}
	F^{\mu \nu}_{~~;\mu} = 0 \label{maxwell}
\end{equation}
for a dipole moment $\mbox{\boldmath $\mu$}=\mu \hat{z}$ we arrive at the 
solution in the Schwarzschild spacetime
\begin{equation}
	A_{\phi} (r,\theta) = f(r) \frac{\mu \sin ^{2} \theta}{r}
\end{equation}
where we have written the expression as the flat-space result weighted
by the multiplicative function:
\begin{equation}
	f(r) = \frac{3r^{3}}{8M^{3}} \left[ \log \eta^2 +
		\frac{2M}{r} \left( 1+ \frac{M}{r} \right) \right]
\end{equation}
\citep[e.g.,][]{was83}.
For the more general rotating case, we multiply the flat space time dependent
vector potential by the same function $f(r)$:
\begin{equation}
	\mbox{\boldmath $A$} = f(r) \left\{ \frac{ \mbox{\boldmath $\mu$}(t) 
		\mbox{\boldmath $\times$ } \mbox{\boldmath $r$} }{r^{3}} +
		\frac{ \dot { \mbox{\boldmath $\mu$}}(t) \mbox{\boldmath $\times$ } 
		\mbox{\boldmath $r$} }{r^{2}} \right\}
\end{equation}
where we now have a time dependent dipole moment,
$\mbox{\boldmath $\mu$}(t) = \mu( -\sin \alpha \cos \Omega_* (t-r), 
-\sin \alpha \sin \Omega_* (t-r), \cos \alpha )$, 
tilted with respect to the rotational ($z$) axis by an angle $\alpha$.
We use 
\{{\boldmath $e_t$},{\boldmath $e_r$},{\boldmath $e_\theta$},{\boldmath $e_\phi$}\} 
to denote the coordinate basis vectors and
\{{\boldmath $e_{\hat{t}}$},{\boldmath $e_{\hat{r}}$},{\boldmath $e_{\hat{\theta}}$},{\boldmath $e_{\hat{\phi}}$}\} 
for the orthonormal basis vectors.  By taking the dot product of the
vector potential with the coordinate basis vectors, we obtain the
vector potential components $(A_t, A_r, A_\theta, A_\phi)$.  Using
\begin{equation}
  \begin{array}{ccccl}
	B_{ \hat{r} } & = & F_{ \hat{\phi} \hat{\theta} } & = &
		\frac{1}{r^{2}\sin \theta } F_{ \phi \theta } \\
	B_{ \hat{\theta} } & = & F_{ \hat{r} \hat{\phi} } & = &
		\frac{ \eta }{r \sin \theta } F_{r \phi } \\
	B_{ \hat{\phi} } & = & F_{ \hat{ \theta } \hat{r} } & = & 
		\frac{ \eta }{r} F_{ \theta r } \\
  \end{array}
\end{equation}
\citep[c.f.][]{mus97}
and the definition of Maxwell's Stress Tensor (eqn.~\ref{maxstr}), 
we derive an expression for the field:
\begin{equation}
   \begin{array}{ccl}
	B_{\hat{r}}      & = & \frac{6 \mu}{(2M)^{3}}
		\chi \left[ -\cos \alpha \cos \theta \right. \\
		&&\left. + \sin \alpha \sin \theta
		\left( \cos \tilde{\phi}  + r \Omega_* \sin \tilde{\phi}
		\right) \right] \\

	B_{\hat{\theta}} & = & \frac{3 \mu \eta}{(2M)^{3}} \left[
		r^{2} \Omega_*^{2} \chi \cos \theta \cos \tilde{\phi} 
		\sin \alpha \right. \\
		&& \left. \mbox{} + 
		2 \left( \log \eta^{2} + \frac{x(1-x/2)}{\eta^{2}} \right)
		\left\{ \cos \alpha \sin \theta \right. \right.\\
		&& \left. \left. + \cos \theta \sin \alpha
		\left( \cos \tilde{\phi}  + r \Omega_* \sin \tilde{\phi}
		\right) \right\} \right] \\

	B_{\hat{\phi}}   & = & \frac{3 \mu \eta \sin \alpha}{(2M)^{3}}
		\left[ - r^{2} \Omega_*^{2} \chi \sin \tilde{\phi} \right. \\
		&& \left. +
		\left( \frac{x(2-x)}{\eta^{2}} + 2\log \eta^{2} \right)
		\left( r \Omega_* \cos \tilde{\phi} - \sin \tilde{\phi}
		\right) \right]
   \end{array}
\end{equation}
with $x=2M/r$, $\chi = \log \eta^{2} + x(1+x/2)$, and
$\tilde{\phi} = \phi - \Omega_* (t-r)$.

The above expressions recover the familiar results in the appropriate
limits:  for $M \rightarrow 0$ we obtain the flat space retarded fields
and for $\Omega_* \rightarrow 0$ we get the Schwarzschild expression
for a dipole tilted with an angle $\alpha$ with respect to the $z$
axis.  Checking Maxwell's Equations (\ref{maxwell}), we find
that two are exactly satisfied ($\nu=1,4$).  The other two are nonzero
with leading terms of order $\mathcal{O}[(\Omega_*r/c)^2(GM/rc^2)]$.

In Figure~\ref{magfld}, we display the field line differences induced
at small $r$ in curved space coordinates.   We have drawn field lines
from the stellar surface in the {\boldmath $\Omega_*$}-{\boldmath $\mu$} plane.
Note that the field lines are pulled closer to the star by curved
space distortions.  The open field lines are nearly radial and thus
display smaller curved space distortions.

\section{Magnetospheric Scattering}

Surface photons ({\it e.g.} from a thermal polar cap) must propagate
through the magnetosphere surrounding the pulsar.  Photons having an
energy in the correct range pass through a local cyclotron resonance
in the magnetosphere. For each photon energy, this defines a
re-emission shell in the closed zone (Figure~\ref{magfld}) in which
strong scattering off the $e^\pm$ re-directs flux along the local
magnetic field line direction.  For millisecond pulsars with $B \sim
10^8-10^9$~gauss, closed zone resonance typically occurs for photons
of energy $1-10$eV.  With typical surface temperatures of a heated
polar cap at $10^6$K, this radiation falls in the Rayleigh-Jeans part
of the blackbody spectrum.

In our Monte Carlo code, each photon trajectory represents a range of
emitted photon energies.  For each photon energy $E_\gamma$, we find
the local cyclotron resonance position along the curved space path as
the photon redshifts to lower energy.  For rapidly rotating stars, we
must include Doppler shifts by boosting the redshifted photon energy
into the local plasma rest frame, checking for resonance, re-emitting
in this locally co-rotating frame, and then boosting back to the lab
frame where we continue to integrate the photon trajectory.  We do not
scatter at open zone resonance positions, since these are assumed to
have no stationary plasma.

In principle, photons can undergo multiple scatterings in the
magnetosphere, for sufficiently large optical depth $\tau$.  Following
\citet{raj97}, we estimate $\tau$ assuming a stationary, flat space
dipolar magnetic field.  The cyclotron energy as a function of
distance from the star is
\begin{equation}
E_{\rm c} = 11.6 B^{*}_{12} (1+3\cos^2 \theta_{\rm B})^{1/2}(R_*/r)^3 {\rm keV}
\end{equation}
where $ B^{*}_{12} $ is the equatorial surface dipole field strength
in units of $10^{12}$~gauss, $\theta_{\rm B}$ is the polar angle
between the dipole moment and the direction of the local magnetic
field, and $R_*$ is the neutron star radius.  The resonant cyclotron
cross section is
\begin{equation}
\sigma_{\rm res} = \frac{\alpha_{\rm F} h^2}{m_{\rm e}}|e_{-}|^2 \delta 
  (E_{\gamma}-E_{\rm c}),
\end{equation}
where $e_-$ contains the dependence
on the photon direction and polarization and $\alpha_{\rm F}$ is the
fine structure constant \citep{mes92}.  We assume a co-rotation
(Goldreich-Julian) charge density $n_{\rm GJ} = 7 \times 10^{13}
B_{12}^{z}/P_{*}({\rm ms})~{\rm cm^{-3}}$ \citep{gol69}.  For rapid
magnetospheric pair production the total $e^\pm$ density may, of
course, be higher.  Integrating the resonant cross section through the
co-rotation charge density we get the optical depth for scattering
\begin{eqnarray}
\tau_{\rm GJ}(E_{\gamma})\!\!\!\!\!&=&\!\!\!\!\!\int_{R_*}^{\infty} n(r) 
	\sigma_{\rm res} {\rm d}r \nonumber \\
&=&\!\!\!\!\!\frac{2.3}{P_{*}({\rm ms})} B^{*1/3}_{12}
      \left( \frac{3 {\rm eV}}{E_{\gamma}} \right)^{1/3} \frac{B_z}{B}
      \frac{R_*}{10 {\rm km}} \label{tau}
\end{eqnarray}
where we have taken $\langle |e_{-}|^2 \rangle = 1/3$.

For concreteness, we adopt fiducial parameters for a millisecond
pulsar ($P_*=3$ms, $R_*=10$km, and $B_*=10^9$~gauss); these give a
characteristic optical depth of $\tau = 7.7 \times 10^{-2} B_z/B$ for
$E_\gamma =3$eV.  For polar cap emission, the average optical depth is
typically somewhat smaller as significant flux travels through the
scattering-free open zone.  Thus, for co-rotation charge densities, we
neglect multiple scatterings.

\begin{figure}[htb]
\epsscale{1}
\plotone{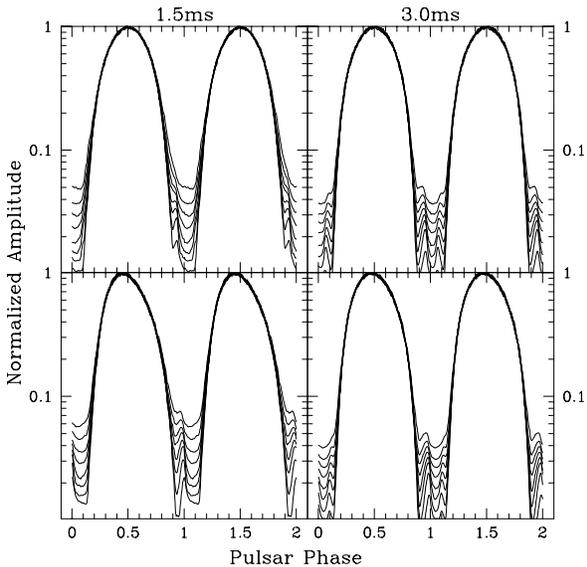}
\caption{Normalized MSP light curves for a single thermal cap and
viewing angle $\zeta= 115\arcdeg$, including closed zone resonant scattering.
\emph{Left} -- $P_*=1.5$ms.
\emph{Right} -- $P_*=3.0$ms.
\emph{Upper panels} -- simple Schwarzschild metric propagations.
\emph{Lower panels} -- propagations include Doppler shifts and time delays.
Curves are plotted for $E_\gamma$ = $1.6$eV (top) to $16$eV (bottom).
\label{lcurv_scat}}
\end{figure}

In Figure~\ref{lcurv_scat}, we plot light curves for thermal emission
from a single polar cap, including closed zone scattering.  The
asymmetry in the scattered flux for the simple Schwarzschild
propagations is due to the field `sweep-back'.  Doppler boosts,
aberration, and time delay effects are quite important,  producing
asymmetry in the direct flux and further phase shifting the scattered
components.  The `inter-pulse' peaks in the scattered flux are due to
radiation scattered back past the star and gravitationally focused to
produce peaked pulse components.  The strong energy dependence of the
location of the scattering screen can been seen in the shifts of this
pulse component.  Note that the scattering amplitude decreases with
$E_\gamma$ as is apparent from equation~(\ref{tau}).  The energy
dependence of the phase shifts is larger for rapid rotators (small
$P_*$), as the high altitude scattering acquires additional aberration
and Doppler boosting.  Formally, these scattered radiation components
provide a precision probe of the near star gravitational field, with
field geometry and other systematics calibrated through the energy
dependence of the component phase.  In principle, precision tests of
strong gravity are possible.  We have run the code with Kerr metric
propagations to check the detectability of these effects.  At
$P_*=1.5$ms we find frame dragging shifts the positions of the
scattered photon peaks (Figure~\ref{lcurv_scat}) by $ \Delta \phi \sim
0.01$.  We conclude that unless neutron stars are very compact at
short periods, or unless closed zone plasmas substantially exceed the
co-rotation density, frame dragging effects will be quite difficult to
measure in scattered photons from surface thermal emission.

\section{Outer Gap Emission}

As a straightforward extension of these models, we consider the
effects of spacetime curvature on the light curves from outer gap
emission, updating the work of \citet{rom95} to include curved space
propagations.  The outer gap region follows the edge of the closed
zone extending from the `null charge surface' (where $n_{\rm GJ}$
changes sign, {\it i.e.} $B_z=0$) to $r_{\rm LC}$.  In outer gap
models this region remains charge-starved, large potential drops
develop, particles accelerate, and non-thermal radiation is produced
tangent to the local $B$ \citep[e.g.,][]{che86, rom96}.  For MSP, this
emission zone comes very close to the neutron star surface where
spacetime curvature becomes important.  

To find the null charge boundary, we make a bicubic spline of the
closed zone surface and step out along the grid until $B_{\rm z}=0$.
We emit tangentially from the `last closed' field line surface with an
emissivity $F \propto r^{-3/2}$.  As described in \citet{rom95}
aberration, boosts, and time delays are all essential for any outer gap
light curve computation.  Accordingly, here we check the additional 
distortions induced by curved space effects, as might be important for
millisecond pulsars.

\begin{figure}[htb]
\epsscale{.74}
\rotatebox{270}{\plotone{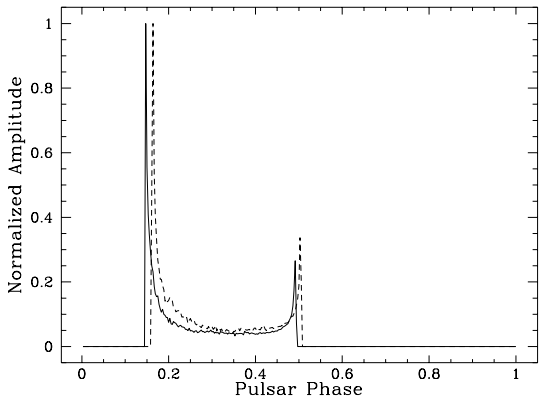}}
\caption{ Light curves for outer gap emission. 
$P_*=1.5$ms, $\alpha = 65\arcdeg$, $\zeta= 115\arcdeg$.  
The magnetic axis (polar cap) is at phase 0.5. Schwarzschild light
curves (solid line) differ from flat space curves (dashed line) in 
both pulse phase and width.
\label{lcurv_open}}
\end{figure}

In Figure~\ref{lcurv_open}, we show sample light curves for outer gap
emission.  The phase of the magnetic axis may be inferred from the
structure of the radio polarization data and so the relative phasing
of the high energy (outer gap) pulse is significant.  Of course even
classical pulsar radio emission, if arising from more than a few
$R_*$, will show similar phase shifts, so comparison of light curve
phases must account for the altitude of the radio emission zone.  The
effect of neglecting Schwarzschild propagations is to both compress
the pulse and shift it to later phase.  For our fiducial parameters
the first pulse component arises at high altitude.  The second pulse
arises from small $r$ where the cap field lines (and photon
directions) are nearly radial, allowing only small gravitational
bending.  The curved space differences are thus dominated by the phase
delays induced by the retardation in the strong near surface field.
The curved space pulse expansion (relative to flat space) can
similarly be attributed to gravitational time delay, since the first
peak arises at larger $r$ where these effects are small.

\section{Conclusions and Observational Prospects}

We have improved the modeling of light curves of rapidly rotating
neutron stars (MSP) by extending the description of the surrounding
vacuum dipole magnetosphere to include curved space and rapid rotation
effects up to corrections of order
$\mathcal{O}[(\Omega_*r/c)^2(GM/rc^2)]$.  These effects produce
substantial changes in the pulse components introduced by
magnetospheric resonant scattering (for thermal surface emission) or
direct emission from the rotating magnetosphere (high altitude gap
radiation).  We have developed a Monte Carlo code to compute light
curves illustrating these effects including the energy dependence of
the scattered photon pulse shapes.  Higher order effects from frame
dragging have also been computed, but these are likely too subtle to
be discernible in magnetospheric pulse components unless the S/N is
very high.  Scattering perturbations, though small, will be greatly
enhanced if the magnetosphere can support a plasma with $\tau >
\tau_{\rm GJ}$.  In this paper, we have only considered the minimum
possible scattering perturbation.  With new detector technologies
\citep[e.g.][]{rom99}, we should see a marked increase in the quality
of optical light curves.  With enhanced sensitivity and spectral
resolution (and possibly higher scattering $\tau$), light curve
perturbations due to rapid rotation effects may be promoted to
important probes of pulsar physics.

\acknowledgments 

This work was supported in part by NASA grant NAG5-3263. Roger W. Romani
is a Cottrell Scholar of Research Corporation.

\end{document}